\documentstyle{aipproc}
\begin{document}
\title{Perturbative superluminal censorship 
       and the null energy condition}
\author{Matt Visser,$^{\dag}$ 
        Bruce Bassett,$^{\P,*}$ and 
        Stefano Liberati$^{\S,\ddagger}$}
\address{$^{\dag}$Physics Department, 
         Washington University,
         Saint Louis, Missouri 63130-4899, USA\\
         $^{\P,\S}$International School for Advanced Studies (SISSA),
         Via Beirut 2--4, 34014 Trieste, Italy\\
         ${}^*$ Department of Theoretical Physics, University of Oxford, 1
         Keble Road, OX1 3NP, UK\\
	 $^{\ddagger}$Istituto Nazionale di Fisica Nucleare (INFN), 
	 sezione di Trieste, Italy}

\maketitle

\begin{abstract}
We argue that ``effective'' superluminal travel, potentially caused by
the tipping over of light cones in Einstein gravity, is always
associated with violations of the null energy condition (NEC). This is
most easily seen by working perturbatively around Minkowski spacetime,
where we use linearized Einstein gravity to show that the NEC forces
the light cones to contract (narrow). Given the NEC, the Shapiro time
delay in any weak gravitational field is always a delay relative to
the Minkowski background, and never an advance. Furthermore, any
object travelling within the lightcones of the weak gravitational
field is similarly delayed with respect to the minimum traversal time
possible in the background Minkowski geometry. 
\end{abstract}

\section*{Introduction}
\def\Lienard{Li\'enard}
\def\Astrofisica{Astrof\'\i{}sica}
\def\Fisica{F\'\i{}sica}

The relationship between the causal aspects of spacetime and the
stress-energy of the matter that generates the geometry is a deep and
subtle one. In this note, which is a simplified presentation based on
our earlier work~\cite{Non-perturbative}, we shall focus in somewhat
more detail on the perturbative investigation of the connection
between the null energy condition (NEC) and the light-cone
structure. We shall demonstrate that in linearized gravity the NEC
always forces the light cones to contract (narrow): Thus the validity
of the NEC for ordinary matter implies that in weak gravitational
fields the Shapiro time delay is always a delay rather than an
advance.

This simple observation has implications for the physics of
(effective) faster-than-light (FTL) travel via ``warp drive''. It is
well established, via a number of rigorous theorems, that any
possibility of effective FTL travel via traversable wormholes
necessarily involves NEC
violations~\cite{Morris-Thorne,MTY,Visser,HV}.  On the other hand, for
effective FTL travel via warp drive (for example, via the Alcubierre
warp bubble~\cite{Alcubierre}, or the Krasnikov FTL
hyper-tube~\cite{Krasnikov}) NEC violations are observed in specific
examples but it is difficult to prove a really general theorem
guaranteeing that FTL travel implies NEC
violations~\cite{Non-perturbative}. Part of the problem arises in even
defining what we mean by FTL, and recent progress in this regard is
reported in~\cite{Non-perturbative,Olum}.

In this note we shall (for pedagogical reasons) restrict attention to
weak gravitational fields and work perturbatively around flat
Minkowski spacetime.  One advantage of doing so is that the background
Minkowski spacetime provides an unambiguous definition of FTL
travel. A second advantage is that the linearized Einstein equations
are simply (if formally) solved via the gravitational
\Lienard--Wiechert potentials. The resulting expression for the metric
perturbation provides information about the manner in which light
cones are perturbed.

\section*{Linearized gravity}

For a weak gravitational field, linearized around flat Minkowski
spacetime, we can in the usual fashion write the metric
as~\cite{Visser,MTW,Wald}
\begin{equation}
g_{\mu\nu} = \eta_{\mu\nu} + h_{\mu\nu},
\end{equation}
with $h_{\mu\nu} \ll 1$. Then adopting the Hilbert--Lorentz gauge
({\em aka} Einstein gauge, harmonic gauge, de Donder gauge, Fock
gauge)
\begin{equation}
\partial_\nu \left[ h^{\mu\nu} - {1\over2} \eta^{\mu\nu} h \right] = 0,
\end{equation}
the linearized Einstein equations are~\cite{Visser,MTW,Wald}
\begin{equation}
\Delta h_{\mu\nu} = -16\pi G 
\left[ T_{\mu\nu} - {1\over2} \eta_{\mu\nu} T \right].
\end{equation}
This has the formal solution~\cite{Visser,MTW,Wald}
\begin{equation}
h_{\mu\nu}(\vec x,t) = 16\pi G \int d^3y \;
{ \left[
T_{\mu\nu}(\vec y,\tilde t) - 
{1\over2} \eta_{\mu\nu} T(\vec y,\tilde t) 
\right]
\over
|\vec x-\vec y|},
\end{equation}
where $\tilde t$ is the retarded time $\tilde t = t - |\vec x-\vec
y|$. These are the gravitational analog of the \Lienard--Wiechert
potentials of ordinary electromagnetism, and the integral has support
on the unperturbed backward light cone from the point $\vec x$. 

In writing down this formal solution we have tacitly assumed that
there is no incoming gravitational radiation. We have also assumed
that the {\em global geometry} of spacetime is approximately
Minkowski, a somewhat more stringent condition than merely assuming
that the metric is {\em locally} approximately Minkowski. Finally note
that the fact that we have been able to completely gauge-fix Einstein
gravity in a canonical manner is essential to argument. That we can
locally gauge-fix to the Hilbert--Lorentz gauge is automatic. By the
assumption of asymptotic flatness implicit in linearized Einstein
gravity, we can apply this gauge at spatial infinity where the only
remaining ambiguity, after we have excluded gravitational radiation,
is that of the Poincare group. (That is: Solutions of the
Hilbert--Lorentz gauge condition, which can be rewritten as $\nabla^2
x^\mu = 0$, are under these conditions unique up to Poincare
transformations.) We now extend the gauge condition inward to cover
the entire spacetime, the only obstructions to doing so globally
coming from black holes or wormholes, which are excluded by
definition. Thus adopting the Hilbert--Lorentz gauge in linearized
gravity allows us to assign a {\em canonical} flat Minkowski metric to
the entire spacetime, and it is the existence of this canonical flat
metric that permits us to make the comparisons (between two different
metrics on the same spacetime) that are at the heart of the argument
that follows.

Now consider a vector $k^\mu$ which we take to be a null vector of the
{\em unperturbed\,} Minkowski spacetime
\begin{equation}
\eta_{\mu\nu} \; k^\mu k^\nu = 0.
\end{equation}
In terms of the full perturbed geometry this vector has a norm
\begin{eqnarray}
||k||^2 &\equiv&
g_{\mu\nu} \; k^\mu k^\nu 
\\
&=&
h_{\mu\nu} \; k^\mu k^\nu
\\
&=& 16\pi G \int d^3y \;
{ 
T_{\mu\nu}(\vec y,\tilde t)  \; k^\mu k^\nu
\over
|\vec x-\vec y|}.
\end{eqnarray}
Now assume the NEC
\begin{equation}
T_{\mu\nu} \; k^\mu k^\nu \geq 0, 
\end{equation}
and note that the kernel $|\vec x-\vec y|^{-1}$ is positive definite.
Using the fact that the integral of a everywhere positive integrand is
also positive, we deduce $g_{\mu\nu} \; k^\mu k^\nu \geq 0$.  Barring
degenerate cases, such as a completely empty spacetime, the integrand
will be positive definite so that
\begin{equation}
g_{\mu\nu} \; k^\mu k^\nu > 0.
\end{equation}
That is, a vector that is null in the Minkowski metric will be
spacelike in the full perturbed metric. Thus the null cone of the
perturbed metric must everywhere lie inside the null cone of the
unperturbed Minkowski metric.

Because the light cones contract, the {\em coordinate} speed of light
must everywhere decrease. (Not the {\em physical} speed of light as
measured by local observers, as always in Einstein gravity, that is of
course a constant.) This does however mean that the time required for
a light ray to get from one spatial point to another must always
increase compared to the time required in flat Minkowski space. This
is the well-known Shapiro time delay, and we see two important points:
(1) to even define the delay (delay with respect to what?) we need to
use the flat Minkowski metric as a background, (2) the fact that in
the solar system it is always a delay, never an advance, is due to the
fact that everyday bulk matter satisfies the NEC.

(We mention in passing that the strong energy condition [SEC] provides
a somewhat stronger result: If the SEC holds then the proper time
interval between any two timelike separated events in the presence of
the gravitational field is always larger than the proper time interval
between these two events as measured in the background Minkowski
spacetime.)

Now subtle quantum-based violations of the NEC are known to
occur~\cite{Visser:ANEC}, but they are always small and are in fact
tightly constrained by the Ford--Roman quantum
inequalities~\cite{Ford-Roman,Pfenning-Ford}. There are also {\em
classical} NEC violations that arise from non-minimally coupled scalar
fields~\cite{Flanagan-Wald}, but these NEC violations require
Planck-scale expectation values for the scalar field.  NEC violations
are never appreciable in a solar system or galactic setting. (SEC
violations are on the other hand relatively common. For example:
cosmological inflation, classical massive scalar fields, etc.)

{From} the point of view of warp drive physics, this analysis is
complementary to that of~\cite{Olum}, (and also to the comments by
Coule~\cite{Coule}, regarding energy condition violations and
``opening out'' the light cones).  Though the present analysis is
perturbative around Minkowski space, it has the advantage of
establishing a direct and immediate {\em physical} connection between
FTL travel and NEC violations. Generalizing this result beyond the
weak field perturbative regime is somewhat
tricky~\cite{Non-perturbative}, and we have addressed this issue
elsewhere. To even define effective FTL one will need to compare two
metrics. (Just to be able to ask the question ``FTL with respect to
what?'').

Even if we simply work perturbatively around a general metric, instead
of perturbatively around the Minkowski metric, the complications are
immense: (1) the Laplacian in the linearized gravitational equations
must be replaced by the Lichnerowicz operator; (2) the Green function
for the Lichnerowicz operator need no longer be concentrated {\em on}
the past light cone [physically, there can be back-scattering from the
background gravitational field, and so the Green function can have
additional support from {\em within} the backward light cone]; and (3)
the Green function need no longer be positive definite.

For example, even for perturbations around a
Friedman--Robertson--Walker (FRW) cosmology, the analysis is not
easy~\cite{FRW-Ford}. Because linearized gravity is {\em not}
conformally coupled to the background the full history of the
spacetime back to the Big Bang must be specified to derive the Green
function. From the astrophysical literature concerning gravitational
lensing it is known that {\em voids} (as opposed to over-densities)
can sometimes lead to a Shapiro time {\em advance}
\cite{Advance1,Advance2,Advance3}. This is not in conflict with the
present analysis and is not evidence for astrophysical NEC
violations. Rather, because those calculations compare a inhomogeneous
universe with a void to a homogeneous FRW universe, the existence of a
time advance is related to a suppression of the density below that of
the homogeneous FRW cosmology.  The local speed of light is determined
by the local gravitational potential relative to the FRW
background. Voids cause an increase of the speed of photons relative
to the homogeneous background. The total time delay along a particular
geodesic is, however, affected by two factors: the gravitational
potential effect on the speed of propagation and the geometric effect
due to the change in path of the photon (lensing) which may make the
total path length longer. Thus traveling through a void doesn't {\em
necessarily} imply an advance relative to the background geometry.

\section*{Discussion} 

This note argues that any form of FTL travel requires violations of
the NEC.  The perturbative analysis presented here is very useful in
that it demonstrates that it is already extremely difficult to even
get even started: Any perturbation of flat space that exhibits even
the slightest amount of FTL (defined as widening of the light cones)
must violate the NEC. The perturbative analysis also serves to focus
attention on the Shapiro time delay as a diagnostic for FTL, and it is
this feature of the perturbative analysis we have extended elsewhere
to the non-perturbative regime to provide both a non-perturbative
definition of FTL~\cite{Non-perturbative}, and a non-perturbative
theorem regarding superluminal censorship.

\bigskip



\begin{references}
\bibitem[\dag]{e-mail}Electronic mail: visser@kiwi.wustl.edu
\bibitem[\P]{e-mail}Electronic mail: bruce@thphys.ox.ac.uk
\bibitem[\S]{e-mail}Electronic mail: liberati@sissa.it          
\bibitem{Non-perturbative}
M. Visser, B. Bassett, and S. Liberati, 
{\em Superluminal censorship}, gr-qc/9810026.
\bibitem{Morris-Thorne}
M.S. Morris and K.S. Thorne, 
Am. J. Phys. {\bf 56}, 395 (1988).
\bibitem{MTY} 
M.S. Morris, K.S. Thorne, and U. Yurtsever, 
Phys. Rev. Lett, {\bf 61}, 1446 (1988).
\bibitem{Visser}
M. Visser, 
{\em Lorentzian wormholes}, 
(AIP Press, New York, 1995).
\bibitem{HV}
D. Hochberg and M. Visser, 
Phys. Rev. Lett. {\bf 81}, 746 (1998);
Phys. Rev. {\bf D57}, 044021 (1998).
\bibitem{Alcubierre} 
M. Alcubierre, 
Class. Quantum Grav. {\bf 11}, L73 (1994).
\bibitem{Krasnikov} 
S.V. Krasnikov,
Phys. Rev. {\bf D57}, 4760 (1998).
\bibitem{Olum} 
K. Olum, 
Phys. Rev. Lett. {\bf 81}, 3567 (1998)
\bibitem{MTW} 
C.W. Misner, K.S.  Thorne, and J.A. Wheeler,
{\em Gravitation},
(Freeman, San Francisco, 1973).
\bibitem{Wald} 
R.M. Wald, 
{\em General Relativity}, 
(Chicago University Press, 1984).
\bibitem{Visser:ANEC}
M. Visser, 
Phys. Lett. {\bf B349}, 443 (1995);
Phys. Rev. {\bf D54}, 5103 (1996); {\bf D54}, 5116 (1996);
{\bf D54}, 5123 (1996); {\bf D56}, 936 (1997). 
\bibitem{Ford-Roman} 
L.H. Ford and T.A. Roman,
Phys. Rev. {\bf D51}, 4277 (1995);  {\bf D53}, 5496 (1996)
\bibitem{Pfenning-Ford}  
M.J. Pfenning and L.H. Ford, 
Class. Quantum Grav. {\bf 14}, 1743 (1997); 
\bibitem{Flanagan-Wald}
E.E. Flanagan and R.M. Wald, 
Phys. Rev. {\bf 54}, 6233 (1996).
\bibitem{Coule}
D.H. Coule, 
Class. Quantum Grav. {\bf 15}, 2523 (1998).
\bibitem{FRW-Ford}
M.J. Pfenning and L.H. Ford, Phys. Rev. {\bf D55}, 4813, (1997).
\bibitem{Advance1}                           
P. Schneider, J. Ehlers, and E.E. Falco,
{\em Gravitational Lenses}, (Springer--Verlag, Berlin, 1992).
\bibitem{Advance2} 
N. Mustapha, B.A. Bassett, C. Hellaby, and G.F.R. Ellis,  
Class. Quantum Grav. {\bf  15}, 2363 (1998).
\bibitem{Advance3}
G.F.R. Ellis and D.M. Solomons,
Class. Quantum Grav. {\bf 15} 2381 (1998).
\end{references}
\end{document}